\documentclass[12pt]{amsart}
\usepackage{amssymb,latexsym}
\usepackage[dvips,draft]{graphicx}

\begin{document}
\title[Wormholes supported by small amounts of exotic matter]
{More on wormholes supported by small amounts of exotic matter}
\author{Peter K. F. Kuhfittig}
\address{Department of Mathematics\\
Milwaukee School of Engineering\\
Milwaukee, Wisconsin 53202-3109}
\date{\today}

\begin{abstract}
Recent papers by Fewster and Roman have emphasized that wormholes
supported by arbitrarily small amounts of exotic matter will have to
be incredibly fine-tuned if they are to be traversable.  This paper 
discusses a wormhole model that strikes a balance between 
two conflicting requirements, reducing the amount of exotic matter 
and fine-tuning the metric coefficients, ultimately resulting in an 
engineering challenge: one requirement can only be met at the expense
of the other.  The wormhole model is macroscopic and satisfies 
various traversability criteria.
\end{abstract}

\maketitle 

PAC number(s): 04.20.Jb, 04.20.Gz

\phantom{a}

\section{Introduction}
Two recent papers by Fewster and Roman \cite{FR05a, FR05b} discuss 
a variety of problems arising from wormholes using arbitrarily small
amounts of exotic matter, including wormhole models by Visser 
\cite{VKD03} and Kuhfittig \cite{pK99, pK02, pK03}.  It was found that
two of the Kuhfittig models are seriously flawed by not taking into 
account the proper distances in estimating the size of certain 
wormholes.  A corrected version of these models resulted in a wormhole 
that turned out to be traversable but contained an exotic region 
whose proper thickness was much larger than desired for wormhole 
construction \cite{FR05b}.  Some of the other models discussed 
in \cite{FR05a} need to be incredibly fine-tuned, even for small 
throat sizes.

The purpose of this paper is to propose a reasonable balance between 
reducing the proper thickness of the exotic region (the region in which 
the weak energy condition is violated) and the amount of 
fine-tuning required to achieve this reduction.  There are no 
particular restrictions on the throat size.  Various constraints are 
shown to be met, so that the resulting wormhole is traversable for 
humanoid travelers.


\section{The basic model}
Our starting point is the line element \cite{pK03}
\begin{equation}\label{E:line}
   ds^2 =-e^{2\gamma(r)}dt^2+e^{2\alpha(r)}dr^2+r^2(d\theta^2+
      \text{sin}^2\theta\, d\phi^2),
\end{equation}
where $\gamma(r)$ is the redshift function, which must be everywhere
finite.  Our units are taken to be those in which $G=c=1$.
The function $\alpha$ has a vertical asymptote at the throat 
$r=r_0$:
\[ 
   \lim_{r \to r_0+}\alpha(r)=+\infty.
\]
If $e^{2\alpha(r)}=1/(1-b(r)/r)$ [Morris and Thorne 
(MT) \cite{MT88}], then the ``shape function" $b(r)$ can be written 
\begin{equation}\label{E:shape}
    b(r)=r(1-e^{-2\alpha(r)}).
\end{equation}
In the absence of any other conditions the proper radial distance 
$\ell(r)$ to the throat from any point outside, given by
\begin{equation}\label{E:proper1}
   \ell(r)=\int\nolimits_{r_0}^{r}e^{\alpha(r')}dr',
\end{equation}
may diverge as $r\rightarrow r_0$.  One way to avoid this problem is 
to start with the function
\begin{equation}
  \alpha_1(r)=\text{ln}\frac{K}{(r-r_0)^A},\qquad 0<A<1,
\end{equation}
in the vicinity of the throat; $K$ is a constant having the same units
as $(r-r_0)^A$ and will be determined later.  Since this function 
eventually becomes negative, it will have to be joined smoothly to 
some $\alpha_2(r)>0$ which goes to zero as $r\rightarrow \infty$.  
(This construction will be made explicit below.) 

Now observe that for $\alpha_1(r)$
\begin{equation}\label{E:proper2}
   \ell(r)=\int\nolimits_{r_0}^{r}e^{\text{ln}\frac{K}{(r'-r_0)^A}}dr'
    =\int\nolimits_{r_0}^{r} \frac{K}{(r'-r_0)^A}dr',
\end{equation}
which is finite for $0<A<1$.  In fact, $\ell(r_0)=0$.  The redshift function
is assumed to have a similar form in the vicinity of the throat:
\begin{equation}\label{E:redshift1}
   \gamma_1(r)=-\text{ln}\frac{L}{(r-r_2)^B},\qquad 0<B<1,
\end{equation}
where $0<r_2<r_0$ to avoid an event horizon at the throat.  This
function is also subject to modification.
  
Next, recall that the weak energy condition (WEC) requires the 
mass-energy tensor $T_{\alpha\beta}$ to obey 
\[
     T_{\alpha\beta}\mu^{\alpha}\mu^{\beta}\ge0  
\]
for all time-like vectors and, by continuity, all null vectors.
Using the notation in Ref.~ \cite{pK03}, the violation of the weak energy 
condition is given by $\rho-\tau<0$, where
\begin{equation}\label{E:WEC}
  \rho-\tau=\frac{1}{8\pi}\left[\frac{2}{r}e^{-2\alpha(r)}
    \left[\alpha'(r)+\gamma'(r)\right]\right].
\end{equation}
[Sufficiently close to the asymptote, $\alpha'(r)+\gamma'(r)$ is 
clearly negative.]  To satisfy the Ford-Roman constraints, we would 
like the WEC to be satisfied outside some small interval $[r_0,r_1]$.  
To accomplish this, choose $r_1$ and construct $\alpha$ and $\gamma$ 
so that
\[
   |\alpha'_1(r_1)|=|\gamma'_1(r_1)|.
\]
It will be shown presently that, if 
\begin{equation}\label{E:AB}
    B=\frac{r_1-r_2}{r_1-r_0}A,
\end{equation}
then $|\alpha'_1(r)|>|\gamma'_1(r)|$ for $r_0<r<r_1$, and 
$|\alpha'_1(r)|<|\gamma'_1(r)|$ for $r>r_1$.  More precisely,
\[
   \alpha'_1(r)=\frac{-A}{r-r_0}<\frac{-A(r_1-r_2)}{r_1-r_0}
    \frac{1}{r-r_2}=-\gamma'_1(r)
\]
for $r_0<r<r_1$; if $r>r_1$, the sense of the inequality is reversed.  
To see this, observe that equality holds for $r=r_1$.  Now multiply 
both sides by $r-r_2$ and denote the resulting left side by $f(r)$, i.e.,
\[
    f(r)=\frac{-A(r-r_2)}{r-r_0}.
\]
 Since $f'(r)>0$, $f(r)$ is strictly increasing, enough to establish the
inequalities.

\section{Some modifications}\label{S:mod}
As noted earlier, $\alpha_1$ and $\gamma_1$ will eventually 
become negative.  Anticipating this, we can cut these functions off at
some $r=r_3$ and then connect them smoothly to suitable new functions
$\alpha_2$ and $\gamma_2$.
(For physical reasons $r_3$ should be larger than $r_1$.)  Suppose 
$\alpha_2$ has the following form:
\[
    \alpha_2(r)=\frac{C}{r-r_0};
\]
then
\[
    \alpha'_2(r)=-\frac{C}{(r-r_0)^2}.
\]
Since we want $\alpha'_1(r_3)=\alpha_2'(r_3)$, we have
\[
   \alpha'_2(r_3)=-\frac{C}{(r_3-r_0)^2}=\frac{-A}{r_3-r_0}.
\]
 It follows that $C=A(r_3-r_0)$ and
\[
      \alpha_2(r)=\frac{A(r_3-r_0)}{r-r_0}.
\]
Similarly,
\[
    \gamma_2(r)=-\frac{B(r_3-r_0)}{r-r_0}.
\]
Besides having slopes of equal absolute value at $r_3$, we want the
functions to meet, i.e., we want
\[
  \alpha_1(r_3)=\alpha_2(r_3)\,\,\text{and}\,\,\gamma_1(r_3)
       =\gamma_2(r_3).
\]
To this end we must determine $K$ and $L$.  For the first case,
\[
    e^{\text{ln}\frac{K}{(r_3-r_0)^A}}=e^{\frac{A(r_3-r_0)}{r_3-r_0}}
     =e^A;
\]
thus $K=e^A(r_3-r_0)^A$.  In a similar manner, $L=e^B(r_3-r_0)^B$.

Returning to $\alpha_1$ and $\gamma_1$ at $r=r_1$, $A$ and $B$ were
chosen so that $\alpha_1$ and $\gamma_1$ would have the desired
properties.  For $r<r_1$ the form of $\gamma_1$ can safely be altered: 
retain the value of the function as well as the slope at $r=r_1$ but 
for $r<r_1$ continue $\gamma_1$ as a straight line with slope 
$\gamma_1(r_1)$ or, better still, as a curve that is approximately 
linear.   (In particular, $\gamma_1$ need not have 
an asymptote close to $r=r_0$.)  As a result,  $|\gamma_1''|$ is
relatively small, while $\gamma'<|\alpha'|$ on $[r_0,r_1]$, as before.

The purpose of this change is two-fold: to avoid a large time-dilation 
near the throat (discussed at the end of Section \ref{S:other}) 
and to address the issue raised in Ref.~\cite{FR05b}: as long as the 
asymptote is retained at $r=r_2$, the proper distance across the exotic
region is shown to be quite large.


\section{The constraints}\label{S:constraints}
Wormhole solutions allowed by general relativity may be subject to severe
constraints from quantum field theory, particularly the quantum 
inequalities \cite{FR95,FR96}, also discussed in \cite{FR05a}.  Of 
particular interest to us is Eq. (95) in Ref.~\cite{FR96}:
\begin{equation}\label{E:QI}
  \frac{r_m}{r_0}\le\left(\frac{1}{v^2-b_0'}\right)^{1/4}
       \frac{\sqrt{\gamma}}{f}\left(\frac{l_p}{r_0}\right)^{1/2},
\end{equation}
where $r_m$ is the smallest of several length scales, $\gamma=
1/\sqrt{1-(v/c)^2}$, $l_p$ is the Planck length, $f$ is a small
scale factor, and $b_0'=b'(r_0)$. Returning to Eq. ~(\ref{E:shape}), 
the shape function, we have
\[
  b'(r)=\frac{d}{dr}\left[r(1-e^{-2\alpha(r)})\right]=
   \frac{d}{dr}\left[r\left(1-\frac{1}{K^2}(r-r_0)^{2A}\right)\right]=1
\]
when $r=r_0$, provided that $A>\frac{1}{2}$, required to meet the radial
tidal constraint, discussed below.  (Observe that $b_0'$ 
does not even exist unless $A\ge\frac{1}{2}$). For the right-hand 
side of inequality (\ref{E:QI}) to be defined and real, we must have  
$v^2>b_0'$.  Since $b_0'=1$ and $v\le 1$, the inequality is trivially 
satisfied, thereby removing any concerns about the size $r_0$ of the 
radius of the throat. Of course, the quantum inequality (\ref{E:QI}) 
cannot simply be circumvented in this manner --- there exist other 
constraints which may be equally severe: (i) as noted in 
Ref.~\cite{FR05a}, a wormhole for which $b_0'=1$ may be unstable 
and (ii)  the exponent $B$ in $\gamma_1(r)$ is subject to severe 
fine-tuning resulting in part from the tidal constraints 
(MT~\cite{MT88}). 

To check these constraints, we need some of the 
components of the Riemann curvature tensor.  From Ref.~\cite{pK02}
\begin{equation}\label{E:radial1}
  R_{\hat{r}\hat{t}\hat{r}\hat{t}}=e^{-2\alpha_1(r)}
   \left(\gamma''_1(r)-\alpha'_1(r)\gamma'_1(r)
      +\left[\gamma'_1(r)\right]^2\right),
\end{equation}
\begin{equation}\label{E:lateral1}
   R_{\hat{\theta}\hat{t}\hat{\theta}\hat{t}}=\frac{1}{r}
     e^{-2\alpha(r)}\gamma_1'(r),
\end{equation}
and
\begin{equation}\label{E:lateral2}
   R_{\hat{\theta}\hat{r}\hat{\theta}\hat{r}}=\frac{1}{r}
       e^{-2\alpha(r)}\alpha_1'(r).      
\end{equation}
For the radial tidal constraint we have
\[
  \left|R_{\hat{1}'\hat{0}'\hat{1}'\hat{0}'}\right|=
  \left|R_{\hat{r}\hat{t}\hat{r}\hat{t}}\right|\le (10^8\,\text{m})^{-2}.
\]
So from Eq.~(\ref{E:radial1}),
\begin{multline}\label{E:radial2}
  |R_{\hat{r}\hat{t}\hat{r}\hat{t}}|=\left|e^{-2\alpha_1(r)}
   \left(\gamma''_1(r)-\alpha'_1(r)\gamma'_1(r)
      +\left[\gamma'_1(r)\right]^2\right)\right|\\
    =\left|e^{-2\,\text{ln}K/(r-r_0)^A}\left(\frac{-B}{(r-r_2)^2}
       -\frac{-A}{r-r_0}\frac{B}{r-r_2}+\frac{B^2}{(r-r_2)^2}\right)
            \right|\\
  =\left|\frac{(r-r_0)^{2A}}{K^2}\left(-\frac{B}{(r-r_2)^2}+\frac{AB}
     {(r-r_0)(r-r_2)}+\frac{B^2}{(r-r_2)^2}\right)\right|\\
     \le (10^8\,\text{m})^{-2}.
\end{multline}
As long as $A>\frac{1}{2}$, as before, 
$|R_{\hat{r}\hat{t}\hat{r}\hat{t}}|$ is close to zero whenever $r$ is 
close to $r_0$.  
Of particular interest is the constraint at $r=r_1$, discussed next. 

\section{The exotic region}\label{S:exotic}
To help determine the size of the exotic region, we examine the
above constraint at $r=r_1$.  Since 
$|\alpha'_1(r_1)|=|\gamma'_1(r_1)|$, we have
\[
   \frac{A}{r_1-r_0}=\frac{B}{r_1-r_2}.
\]
Substituting in Eq.~(\ref{E:radial2}) and using
\[
   r_1-r_2=\frac{B}{A}(r_1-r_0)
\]
from Eq.~(\ref{E:AB}), we get
\begin{multline}\label{E:lateral3}
  |R_{\hat{r}\hat{t}\hat{r}\hat{t}}|=
   \frac{(r_1-r_0)^{2A}}{K^2}\left(-\frac{A}{(r_1-r_0)}\frac{1}
     {(B/A)(r_1-r_0)}+\frac{2A^2}{(r_1-r_0)^2}\right)\\
  =\frac{(r_1-r_0)^{2A}}{K^2}\frac{A^2(2-1/B)}{(r_1-r_0)^2}=
    (10^8\,\text{m})^{-2},      
\end{multline}
assuming now that the constraint is just met.

Taking the square root of the reciprocals, we get
\[
  K(r_1-r_0)^{1-A}\frac{1}{A\sqrt{2-1/B}}=10^8
\]
or
\begin{equation}
  K(r_1-r_0)^{1-A}=10^8A\sqrt{2-\frac{1}{B}}.
\end{equation}
Returning to Eq.~(\ref{E:proper1}), 
\begin{equation}
   \ell(r)=\int\nolimits_{r_0}^{r} K(r'-r_0)^{-A}dr'
   =\frac{K}{1-A}(r-r_0)^{1-A}.
\end{equation}
(See Fig.~1; the graph is plotted using $A=\frac{1}{2}$, $K=5$,
and $r_0=200\,\text{m}$.)  In particular,
\begin{equation}
   \ell(r_1) =\frac{K}{1-A}(r_1-r_0)^{1-A}
    =\frac{A}{1-A}10^8\sqrt{2-\frac{1}{B}}.
\end{equation}
\begin{figure}[htbp]
\begin{center}
\includegraphics[clip=true, draft=false, bb=0 0 299 289, angle=0, width=3.1in, height=3in, 
   viewport=0 0 296 286]{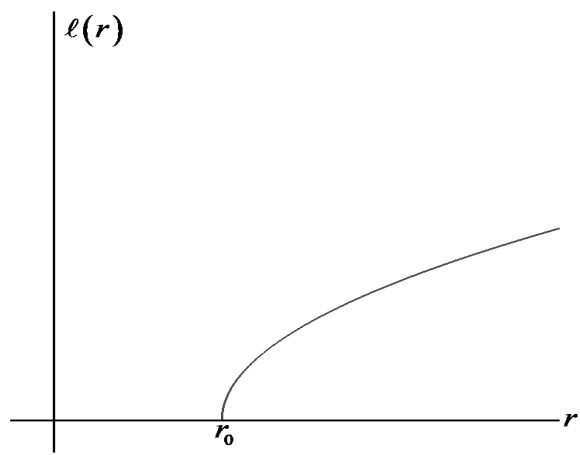}
\end{center}
\caption{\label{fig:figure1}The graph of $\ell(r)=\frac{K}{1-A}(r-r_0)^{1-A}$.}
\end{figure} 
Even though $\ell(r_1)\rightarrow 0$ as $B\rightarrow \frac{1}{2}$, 
considerable fine-tuning is required, as can be seen from Fig.~2
(plotted using $A=\frac{1}{2}$):
$\ell(r_1)$ rises so rapidly that even a small increase in $B$ 
can result in a large increase in $\ell(r_1)$.
\begin{figure}[htbp]
\begin{center}
\includegraphics[clip=true, draft=false, bb=0 0 299 212, angle=0, width=4.25in, height=3.0in, 
   viewport=0 0 296 209]{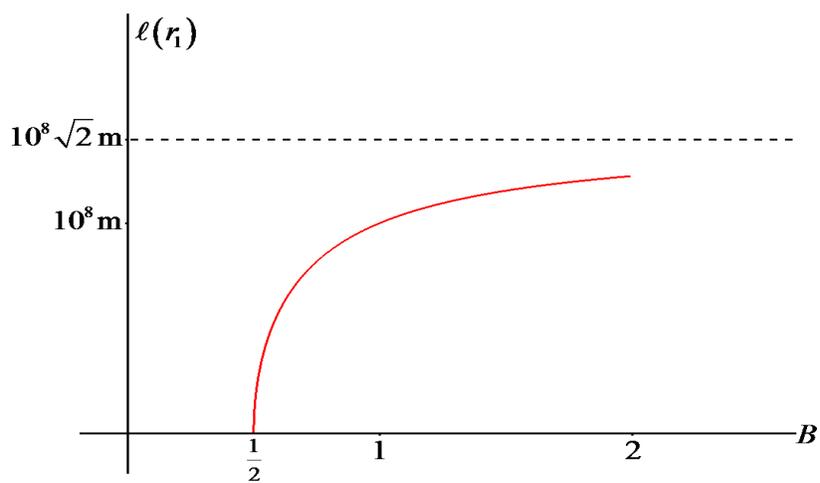}
\end{center}
\caption{\label{fig:figure2}The size of the exotic region as a function of $B$.}
\end{figure} 
Next, let us obtain some numerical estimates: let 
$A=\frac{1}{2}$ again and recall that $K=e^A(r_3-r_0)^A$.  Then 
from Eq.~(\ref{E:lateral3})
\[
  \frac{1}{4}\frac{r_1-r_0}{e(r_3-r_0)}\frac{2-1/B}{(r_1-r_0)^2}=
   \frac{2-1/B}{4e(r_3-r_0)(r_1-r_0)}=(10^8\,\text{m})^{-2}
\]
whence
\[
   B=\left[2-10^{-16}(4e)(r_3-r_0)(r_1-r_0)\right]^{-1}.
\]
(Since $\frac{1}{2}<A<B$, $A$ is also severely restricted.)  In Fig.~3,
$B$ is plotted against log($r_1-r_0$) for four values of $r_3-r_0$
(right to left): 5 m, 50 m, 500 m, and 5000 m.  Observe that the
exponent $B$ remains very close to 1/2 for any reasonable value of
$r_1-r_0$ and is virtually independent of $r_3-r_0$.

\begin{figure}[htbp]
\begin{center}
\includegraphics[clip=true, draft=false, bb=0 0 452 227, angle=0, width=5in, height=2.5in, 
   viewport=0 0 449 224]{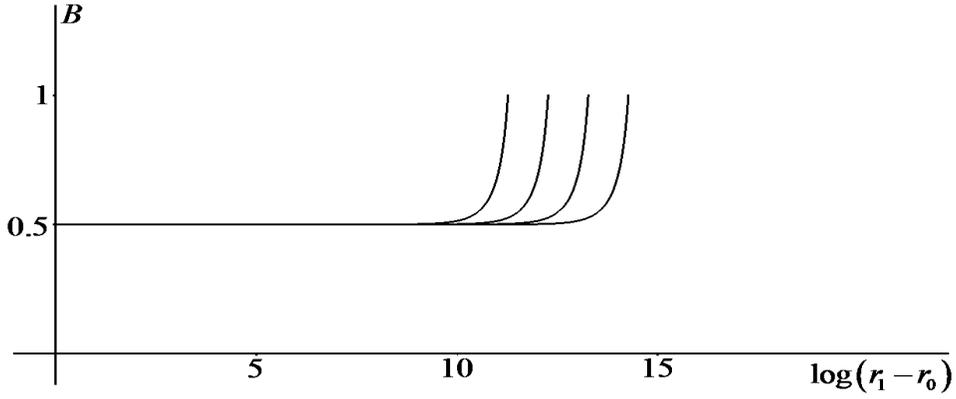}
\end{center}
\caption{\label{fig:figure3}A plot illustrating the dependence of the exponent $B$ on log $(r_1-r_0)$
        for several values of $r_3-r_0$.}
\end{figure} Now consider the following coordinate distances: $r_1-r_0=0.000001$ m
and $r_3-r_0=10$ m. Then 
\[
   B=(2-1.09\times10^{-20})^{-1}.
\]
For the proper distance $\ell(r_1)$ we get
\[
   \ell(0.000001)=\sqrt{1.09\times10^{-20}}(10^8\,\text{m})
   \approx 1\,\text{cm}.
\]
Since Eq. (\ref{E:lateral3}) is really an inequality, the result is,
stricly speaking, $\ell(0.000001)\gtrapprox 1$ cm.  However, the proper 
distance can also be obtained by direct integration, using the 
same values for $K$ and $A$:
\[
   \ell(0.000001)=\int\nolimits_{r_0}^{r_0+0.000001} 
   \frac{K}{(r-r_0)^{A}}dr\approx 1\,\text{cm}.
\]

\emph{Remark:} If $\gamma_1$ is modified for $r<r_1$, as described 
in Section \ref{S:mod}, then the constraint, Eq.~(\ref{E:radial2}), 
is met \emph{a fortiori:} since $(r-r_0)^{2A-1}<(r_1-r_0)^{2A-1}$ 
for any $r<r_1$,
\begin{equation*}
   |R_{\hat{r}\hat{t}\hat{r}\hat{t}}|\approx\frac{1}{K^2}
    \left(0+A(r-r_0)^{2A-1}\gamma'_1(r_1)
    +(r-r_0)^{2A}[\gamma'_1(r_1)]^2\right) 
\end{equation*}
is less than the value of $|R_{\hat{r}\hat{t}\hat{r}\hat{t}}|$
at $r=r_1$.

While the distances chosen are rather arbitrary, the outcome, a proper 
distance of 1 cm, seems promising.  It is also clear that reducing this 
distance any further would require even  more fine-tuning. On the
positive side, without any special restriction on $r_0$, the 
wormhole can be macroscopic.
            
\section{Other constraints}\label{S:other}
It remains to check some of the other constraints in MT \cite{MT88}.
The gradient of the redshift function must satisfy the condition
\begin{equation}
   |\gamma'(r)|\le g_{\oplus}/\left(c^2\sqrt{1-b(r)/r}\right)=
       1.09\times 10^{-16}\,\text{m}^{-1}
\end{equation}
at the stations \cite{MT88}.  If $r$, the distance to the stations,
is chosen to be 225\,000 km and if $r_0 \ll
225\,000$ km, then (since $B\approx1/2$)
\begin{equation*}
  |\gamma_2'(r)|=\frac{B(r_3-r_0)}{(r-r_0)^2}=
    \frac{\frac{1}{2}(10\,\text{m})}
    {(225\,000\,000\,\text{m})^2}
     =1\times10^{-16}\,\text{m}^{-1}.
\end{equation*}
Also, the stations should be far enough away from the throat so that
$b(r)/r\approx 0$, making the space nearly flat.  A similar 
formulation is that $1-b(r)/r=e^{-2\alpha(r)}$ is close to unity. This 
constraint is easily met given that $\alpha_2(r)=A(r_3-r_0)/(r-r_0)$. 

Using Eqs.~(\ref{E:lateral1}) and (\ref{E:lateral2}), the remaining 
tidal constraints can be written \cite{MT88}
\begin{multline*}
  \left|R_{\hat{2}'\hat{0}'\hat{2}'\hat{0}'}\right|
  =\left|R_{\hat{3}'\hat{0}'\hat{3}'\hat{0}'}\right|
  =\gamma^2\left|R_{\hat{\theta}\hat{t}\hat{\theta}\hat{t}}\right|
  +\gamma^2\left(\frac{v}{c}\right)^2\left|
     R_{\hat{\theta}\hat{r}\hat{\theta}\hat{r}}\right|\\
   =\gamma^2\left(\frac{1}{r}e^{-2\alpha(r)}\gamma_1'(r)\right)
   +\gamma^2\left(\frac{v}{c}\right)^2
       \left(\frac{1}{r}\right)e^{-2\alpha(r)}\alpha_1'(r)
     \le (10^8\,\text{m})^{-2};
\end{multline*} 
(here $\gamma^2=1/\left[1-(v/c)^2\right]$.)  The first term is close 
to zero near the throat while the second is merely a constraint on 
the velocity of the traveler.

A final consideration is the time dilation near the throat. Let
$v=d\ell/d\tau$, so that $d\tau=d\ell/v$ (assuming that $\gamma\approx 1$.)  
Since $d\ell=e^{\alpha(r)}dr$ and $d\tau=e^{\gamma(r)}dt$, we have for 
any coordinate interval $\Delta t$:
\begin{equation*}
  \Delta t=\int\nolimits_{t_a}^{t_b}dt=
     \int\nolimits_{\ell_a}^{\ell_b}e^{-\gamma(r)}\frac{d\ell}{v}= 
     \int\nolimits_{r_a}^{r_b}\frac{1}{v}e^{-\gamma(r)}e^{\alpha(r)}
        dr.     
\end{equation*}
Since we are concerned mainly with the vicinity of the throat,
let us consider the interval $[r_0,r_3]$.  If the form of $\gamma_1(r)$
in Eq.~(\ref{E:redshift1}) were retained, the result would be
\begin{equation*}
   \Delta t=\int\nolimits_{r_0}^{r_3}\frac{1}{v}\frac{L}{(r-r_2)^B}
    \frac{K}{(r-r_0)^A}dr.
\end{equation*} 
Since both $A$ and $B$ exceed 1/2, $\Delta t\rightarrow \infty$ as $r_2$
approaches $r_0$.  It was noted in Section~\ref{S:mod}, however, that to
the left of $r_1$, $\gamma_1(r)$ is ``straightened out," that is, 
$\gamma_1(r)$ no longer has an asymptote close to $r=r_2$.  As a result, 
letting $A=\frac{1}{2}$ again,
\begin{equation*}
   \Delta t=\int\nolimits_{r_0}^{r_3}\frac{1}{v}e^{-\gamma_1(r)}
      \frac{K}{(r-r_0)^{1/2}}dr,
\end{equation*}
which is now finite. 
In addition to the proper velocity $v$, the time interval $\Delta t$
evidently depends on $\gamma_1(r)$, as well as the choice of $K$.  But since 
we are dealing with a relatively small distance and hence a small proper 
traversal time, the time dilation is of little significance.  This 
conclusion also holds for $r>r_3$ since $\gamma_2(r)$ and $\alpha_2(r)$
fall off so rapidly.

\section{Summary}
This paper discusses a wormhole model with the following characteristics:
since the derivative of the shape function is unity at the throat, the
quantum inequalities are trivially satisfied.  In place of this requirement
one is faced with a restriction on the spacetime geometry that may be equally
severe.  The severity of this restriction notwithstanding, the model allows
a compromise between shrinking the exotic region and fine-tuning the metric 
coefficients: reducing the amount of exotic matter can only be accomplished
by further fine-tuning.  There is no way to determine \emph{a priori} which
requirement is easier to meet, or whether, being interdependent, either one 
can be met.  One can only hope that in the end this problem reduces to an 
engineering challenge.

No particular restriction is placed on the throat size.  Moreover, various 
traversability criteria are met, resulting in a wormhole that is traversable 
for humanoid travelers.  

\section*{Acknowledgment}
The author would like to thank C.J. Fewster and T.A. Roman for their 
constructive criticisms.

\end{document}